\documentclass[a4paper,11pt]{article}
\usepackage{pos}

\newcommand{\be}{\begin{equation}}
\newcommand{\ee}{\end{equation}}
\newcommand{\bea}{\begin{eqnarray}}
\newcommand{\eea}{\end{eqnarray}}

\newcommand{\la}{\langle}
\newcommand{\ra}{\rangle}

\title{Non-equilibrium dynamics of topological defects \\
  in the 3d O(2) model}
\ShortTitle{Non-equilibrium dynamics of topological defects in
  the 3d O(2) model}

\author*{Edgar L\'{o}pez-Contreras}
\author{Jaime Fabi\'{a}n Nieto Castellanos}
\author{\\El\'{\i}as Natanael Polanco-Eu\'{a}n}
\author{Wolfgang Bietenholz}

\affiliation{Instituto de Ciencias Nucleares,\\
  Universidad Nacional Aut\'{o}noma de M\'{e}xico \\
  A.P.\ 70-543, C.P.\ 04510 Ciudad de M\'{e}xico, Mexico}

\emailAdd{ed\textunderscore lopez@ciencias.unam.mx}
\emailAdd{jafanica@ciencias.unam.mx}
\emailAdd{elias.polanco@correo.nucleares.unam.mx}
\emailAdd{wolbi@nucleares.unam.mx}

\abstract{We present a study of the 3d O(2) non-linear $\sigma$-model on
  the lattice, which exhibits topological defects in the form of vortices.
  They tend to organize into vortex lines that bear close analogies with
  global cosmic strings. Therefore, this model serves as a testbed for
  studying the dynamics of topological defects.
  It undergoes a second order phase transition, hence
  it is appropriate for investigating the Kibble-Zurek mechanism.
  In this regard, we explore the persistence of topological defects
  when the temperature is rapidly reduced from above to below the
  critical temperature; this cooling (or ``quenching'') process takes the
  system out of equilibrium. We probe a wide range of inverse cooling
  rates $\tau_{\rm Q}$ and final temperatures, employing distinct Monte Carlo
  algorithms. The results consistently show that the density of persisting
  topological defects follows a power-law in $\tau_{\rm Q}$, in agreement
  with Zurek's conjecture. On the other hand, at this point our results
  do not confirm Zurek's prediction for the exponent in this
  power-law, but its final test is still under investigation.}

\FullConference{The 40th International Symposium on Lattice Field Theory (Lattice 2023)\\
July 31st - August 4th, 2023\\
Fermi National Accelerator Laboratory\\}

\begin{document}

\maketitle

\section{Kibble mechanism in the Early Universe}

According to established cosmology, the period
$\lesssim 10^{-12}~{\rm s}$ after the Big Bang --- with a
temperature $\gtrsim 160~{\rm GeV}$ --- was dominated
by electroweak interactions. This period ended when the
Higgs field acquired its vacuum expectation value of
$|\la \Phi \ra | > 0$.
The seeds of possible topological defects --- in particular cosmic
strings --- could dated back to that instant.

For their formation, Kibble suggested the following mechanism
\cite{Kibble}: causality did not allow for the complex phases of
$\langle \Phi \rangle$ to be correlated in well separated regions
of the Universe. Cosmic strings may have formed in interfaces
between such regions, when the phase change --- integrated over
a closed loop --- amounted to a non-zero multiple of $2 \pi$.

The Kibble mechanism could have led to a string network
of topological defects.
Later, the cosmic strings could have lost some energy by
gravitational radiation and particle emission, observational
bounds are discussed in Ref.\ \cite{Auclair}. Still, their
topological nature might have stabilized them to this day.

The attempts to observe cosmic strings have not been successful
so far. They seem to be ruled out as seeds for the galaxy
formation: that is incompatible with observations by the
COBE and WMAP satellites of the dominant scale of anisotropies
in the Cosmic Microwave Background \cite{COBE-WMAP}. This does,
however, not imply that cosmic strings do not exist. Recently, the
search for a corresponding signal focuses on gravitational waves,
see {\it e.g.}\ Ref.\ \cite{gravwaves}.

\section{Vortex lines in the 3d O(2) model}

We consider the O(2) non-linear $\sigma$-model on 3-dimensional,
cubic lattices of volumes $V=L^{3}$.
At each lattice site $x$, there is a classical spin
$\vec e_{x} = (\cos \varphi_{x}, \sin \varphi_{x}) \in S^{1}$
and the Hamilton function ${\cal H}$ has the standard form
(in lattice units)
\be
{\cal H}[\vec e \, ] = - J \sum_{\la xy \ra} \vec e_{x} \cdot \vec e_{y} \ ,
\quad Z = \int D\vec e \ \exp(-{\cal H}[\vec e \, ]/T) \ ,
\ee
with a global O(2) symmetry. The sum runs over the nearest neighbor
lattice sites, $Z$ is the partition function and $T$ the temperature
(in units where $k_{\rm B}=1$, and $J=1$).
We always assume periodic boundary conditions.
This model undergoes a second order phase transition at
$T_{\rm c} = 2.2018441(5)$ \cite{Tcrit}. Its universality class is
expected to coincide with the superfluid transition of $^{4}$He,
the so-called $\lambda$-transition, although there is some tension
with the value of the critical exponent $\alpha$ \cite{critexpnu}.

For a given configuration, we assign to each plaquette,
say in the $\mu \nu$ plane, the vorticity
\be  \label{vorticity}
v_{x, \mu \nu} = (\Delta \varphi_{x,x+\hat \mu} +
\Delta \varphi_{x+ \hat \mu, x+\hat \mu + \hat \nu} +
\Delta \varphi_{x+ \hat \mu + \hat \nu, x+\hat \nu} +
\Delta \varphi_{x+ \hat \nu, x}) / 2 \pi
= \left\{ \begin{array}{cc}
 1 & \mbox{vortex} \\
 0 & \mbox{neutral} \\
-1 & \mbox{anti-vortex}
\end{array} \right.
\ee
where $\hat \mu$, $\hat \nu$ are the lattice unit vectors in
$\mu$- and $\nu$-direction,
and $\Delta \varphi_{xy} = \varphi_{y}-\varphi_{x} ~ {\rm mod} ~
2 \pi \in (-\pi, \pi]$ (note the unusual modulo operation).
Ignoring configurations with probability measure zero, each
plaquette carries either a vortex, or an anti-vortex or it is
neutral, as indicated in eq.\ (\ref{vorticity}).

The vortices form {\em vortex lines}, which
connect centers of the lattice unit cubes.
For a vortex and an anti-vortex the connection
of adjacent cubes has opposite
orientation. If the continuation from a cube center is ambiguous,
we make a random choice. In this manner, we obtain vortex lines,
which are closed in practically all cases (taking into account
the periodic boundaries), in remarkable analogy to
global cosmic strings.

\begin{figure}[h!]
\vspace*{-5mm}
\begin{center}
\includegraphics[angle=0,width=.5\linewidth]{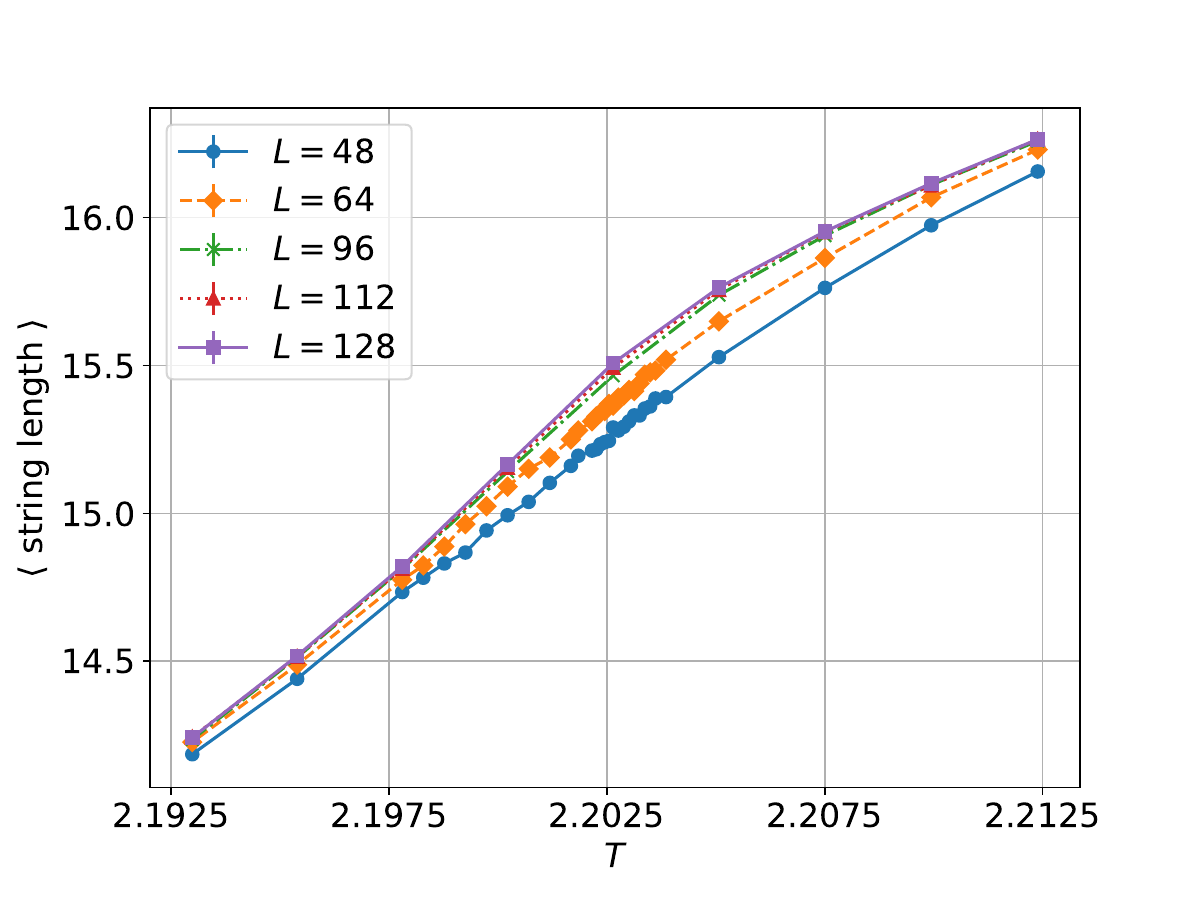}
\hspace*{-3mm}
\includegraphics[angle=0,width=.5\linewidth]{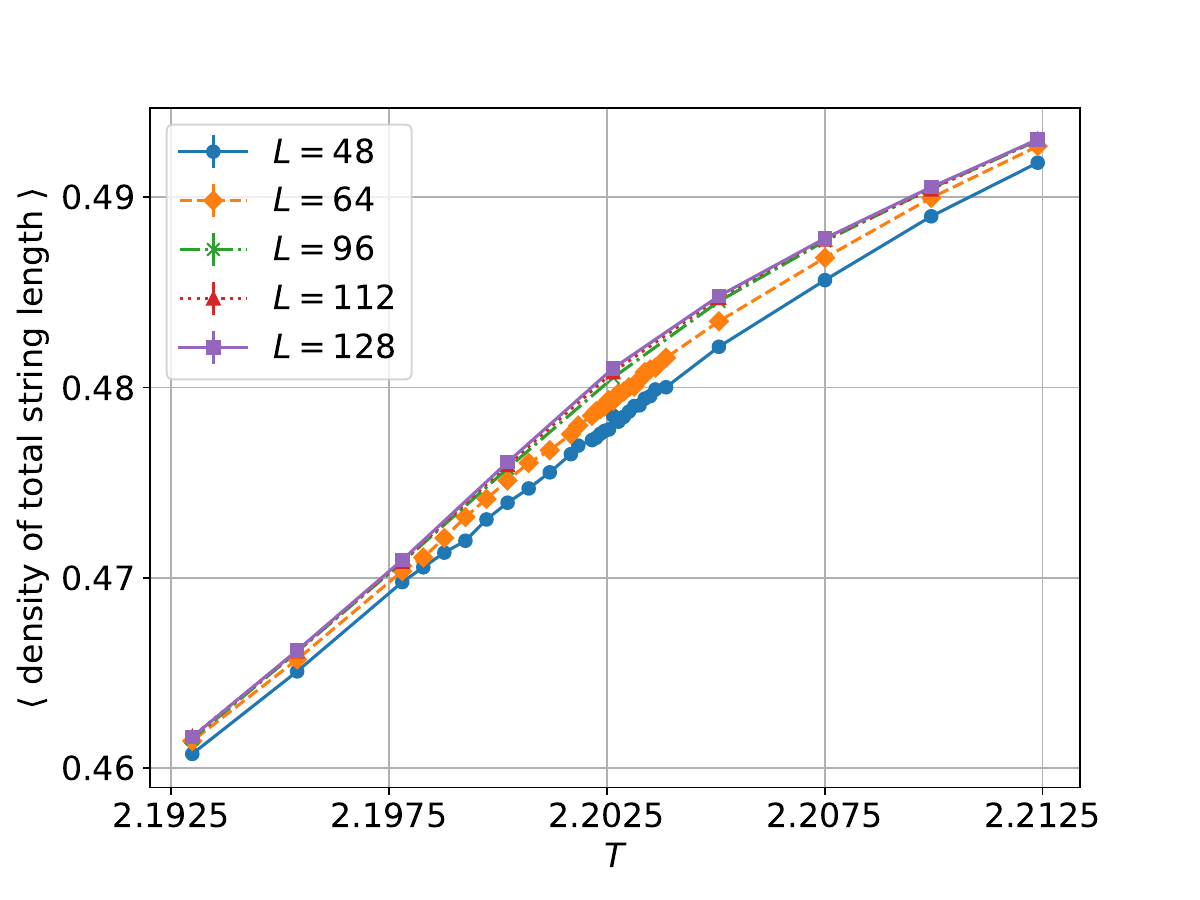}
\end{center}
\vspace*{-5mm}
\caption{Properties of vortex lines in the XY model, in $L^{3}$ lattices,
in analogy to cosmic strings: mean string length (left) and the fractional
volume covered by strings (right), both with modest finite-size effects.}
\vspace*{-5mm}
\label{stringfig}
\end{figure}

Figure \ref{stringfig} shows the mean string length
and the density of the total string length, as functions
of the temperature $T$, in thermal equilibrium. Both
quantities converge well in a volume $V=L^{3}$, when the lattice
size attains $L \approx 100$. A scale for relating them
to the cosmic strings would be set by the correlation length,
which corresponds to $1/m_{\rm Higgs}$.
For a comprehensive study of vortex line properties as a possible
order parameter (which was, however, not exactly confirmed), we refer to
Ref.\ \cite{Finn00}.

\section{Kibble-Zurek mechanism in condensed matter}

Zurek suggested to materialize the Kibble mechanism in condensed
matter systems \cite{Zurek85}, where it is denoted as the
Kibble-Zurek mechanism. Such experiments were successfully
carried out in a non-linear optical system \cite{Ducci} and
with liquid $^{4}$He \cite{LinZurek} .

The Kibble-Zurek mechanism was formulated for systems with a second order
phase transition. At the critical temperature $T_{\rm c}$, the correlation
length $\xi$ and the equilibrium relaxation time $\tau$ diverge as
\be
\xi (\epsilon ) = \frac{C_{\xi}}{|\epsilon |^{\nu}} \ , \quad
\tau (\epsilon ) = \frac{C_{\tau}}{|\epsilon |^{z \nu}} \ ,
\quad {\rm where} \quad \epsilon = \frac{T- T_{\rm c}}{T_{\rm c}}
\ee
is the reduced temperature, $C_{\xi}$ and $C_{\tau}$ are constants,
$\nu$ is a critical exponent (in the usual notation) and $z$ is
the dynamical critical exponent.

In agreement with the literature, we consider a temperature, which
linearly decreases in time $t$. We denote the inverse cooling
(or ``quenching'') rate as $\tau_{\rm Q} > 0$, and formulate the
temperature as
\be  \label{lineartime}
T(t) = T_{\rm c} \Big( 1 - \frac{t}{\tau_{\rm Q}} \Big) \ \to
\ \epsilon (t) = -\frac{t}{\tau_{\rm Q}} \ ; \qquad
t \in [t_{\rm i},\, t_{\rm f}], \quad
t_{\rm i} = -\tau_{\rm Q} \ , \quad 0 < t_{\rm f} \leq \tau_{\rm Q} \ ,
\ee
hence the temperature decreases from its initial value $\, T_{\rm i}
= T(t_{\rm i}) = 2 T_{\rm c}\,$ to $\,T(0) = T_{\rm c}\,$ and finally down to
$\,T_{\rm f} = T(t_{\rm f}) < T_{\rm c}\,$.

\begin{itemize}

\item At an early stage, when we are still far from criticality,
$T$ is clearly larger than $T_{\rm c}$, so
the relaxation time $\tau$ is short, and for
$\tau_{\rm Q} \gg \tau$ the cooling is quasi-adiabatic.

\item As we approach criticality, $T \approx T_{\rm c}$, however,
$\tau$ becomes long and the dynamics departs from equilibrium;
then the system is ``frozen''.

\end{itemize}

Del Campo and Zurek define the transition between the quasi-adiabatic
and quasi-frozen regime at the time $\pm \hat t$, where the elapsed
time before and after crossing criticality coincides with the
relaxation time \cite{Zurek85},
\be
\tau (\hat t) = \hat t \qquad \Rightarrow \qquad
\hat t = (C_{\tau} \tau_{Q}^{z\nu})^{1/(1 + z \nu)} \ .
\ee
At time $\hat t$, the reduced temperature and the correlation
length are given by
\be
\epsilon (\hat t) = - \Big( \frac{C_{\tau}}{\tau_{\rm Q}} \Big)^{1/(1 + z \nu)}
\quad , \qquad
\xi (\hat t) = C_{\xi} \,
\Big( \frac{\tau_{\rm Q}}{C_{\tau}} \Big)^{\nu/(1 + z \nu)}
\ .
\ee
Del Campo and Zurek further estimate the density of topological effects,
which persists at time $\hat t$ (after crossing the critical point) as
\be  \label{rhoend}
\rho = \xi^{\delta -d} = C_{\xi}^{\delta -d} \Big(
\frac{C_{\tau}}{\tau_{\rm Q}} \Big)^{\zeta}
\propto \tau_{\rm Q}^{- \zeta} \ , \qquad
\zeta = \frac{(d - \delta) \nu}{1 + z \nu} \ ,
\ee
where $d$ is the spatial dimension and $\delta$ is the dimension
of the topological defects. Hence this assumption leads to a
power-law with the exponent that we denote as $\zeta$,
$\, \rho \propto 1/ \tau_{Q}^{\zeta}$.

According to Refs.\ \cite{hexman,LinZurek}, experiments with
hexagonal manganites ($R$MnO$_{3}$) are in agreement with this
prediction for large values of $\tau_{\rm Q}$,
with $d=3$, $\delta =1$ and $z=2$.

For the 3d O(2) model, we also have $d=3$ and $\delta =1$,
but $\nu = 0.6717(1)$
\cite{critexpnu}. If the model is simulated with a local-update
algorithm, like Metropolis or heatbath, the dynamical critical
exponent --- with respect to the evolution in Markov time ---
is again known to be $z \approx 2$.
Then Zurek's prediction implies \vspace*{-1mm}
\be  \label{zetaZurek}
\zeta_{\rm Z} \approx 0.57 \ .
\ee

\vspace*{-2mm}
\section{Vortex density after rapid cooling}
\vspace*{-1mm}

For our numerical investigation, we first thermalize the
model at a high temperature of $T=2 T_{\rm c}$; this is
preferably done with the cluster algorithm.

Then we switch to a local update algorithm, Metropolis or heatbath,
and perform lexicographic sweeps of $L^{3}$ spin updates
(here it makes indeed a difference if the updated spins are randomly
selected). In the Metropolis algorithm, the entire
circle $S^{1}$ is considered for a new spin direction
(introducing a limited interval would mean another artificial
ingredient). For the implementation of the heatbath
algorithm, we follow the prescription of Ref.\ \cite{HatNak}.
In this cooling process, the highly efficient cluster algorithm
(which is optimal in equilibrium simulations) does not seem
appropriate, because its evolution in Markov time is too quick
for the cooling process to be well monitored.

After each sweep, the temperature is linearly decreased according
to eq.\ (\ref{lineartime}), until it arrives at $T_{\rm f}$.
We measure the vortex density $\rho$ after each sweep, and
repeat this cooling process 1000 times,
starting every time from a different configuration, which is
thermalized at $2 T_{\rm c}$. For each instant in Markov time,
{\it i.e.}\ after each number of sweeps, we average over the
values of $\rho$ measured in these 1000 processes. We are
particularly interested in the final vortex density $\rho_{\rm f}$;
for rapid cooling, it is well above the equilibrium density at
$T_{\rm f}$ (which vanishes for $T_{\rm f}=0$).

A similar study was performed before in the 2d XY model \cite{JelCug},
where, however, the phase transition is essential (infinite order in
Ehrenfest's scheme), {\it i.e.}\ not of second order as Zurek assumed
when formulating his conjecture.

\begin{figure}[h!]
\vspace*{-5mm}
\begin{center}
\includegraphics[angle=0,width=.5\linewidth]{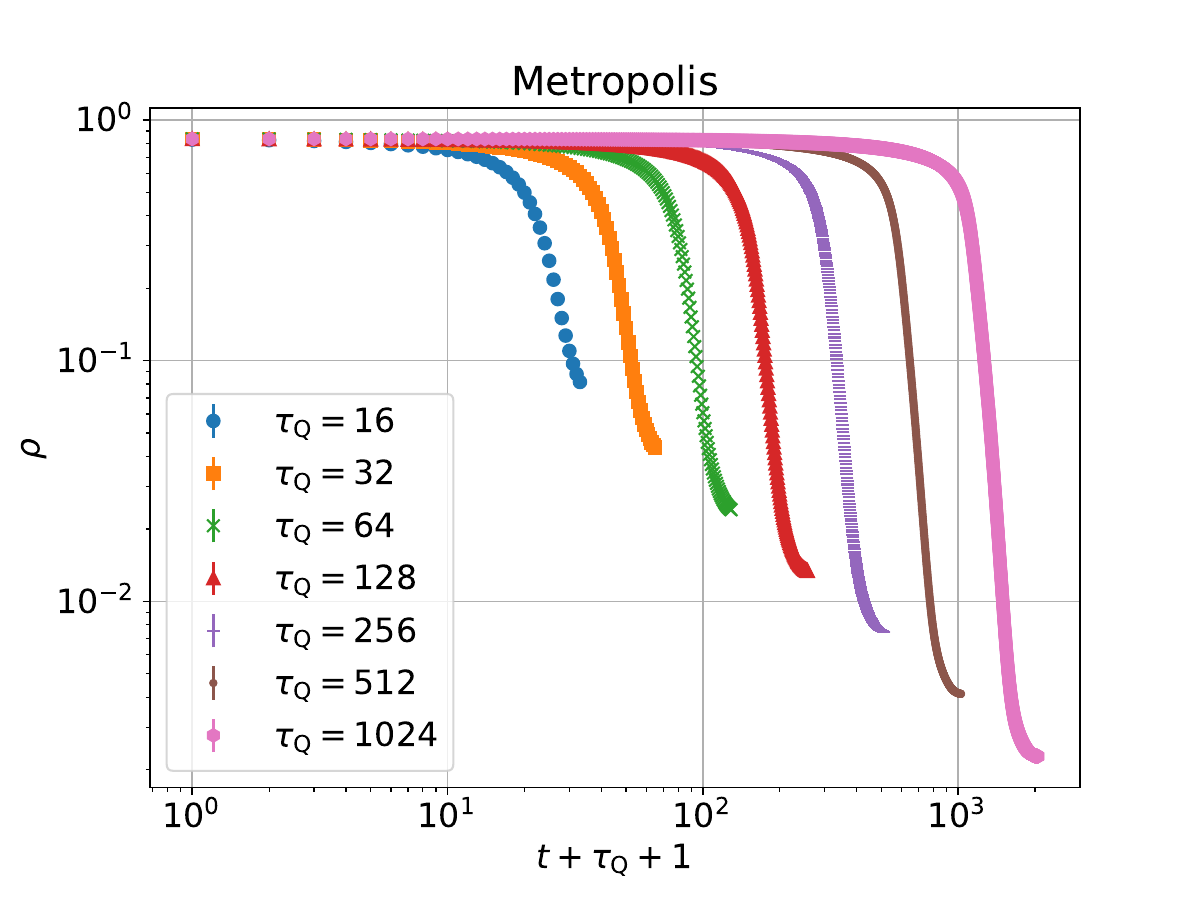}
\hspace*{-3mm}
\includegraphics[angle=0,width=.5\linewidth]{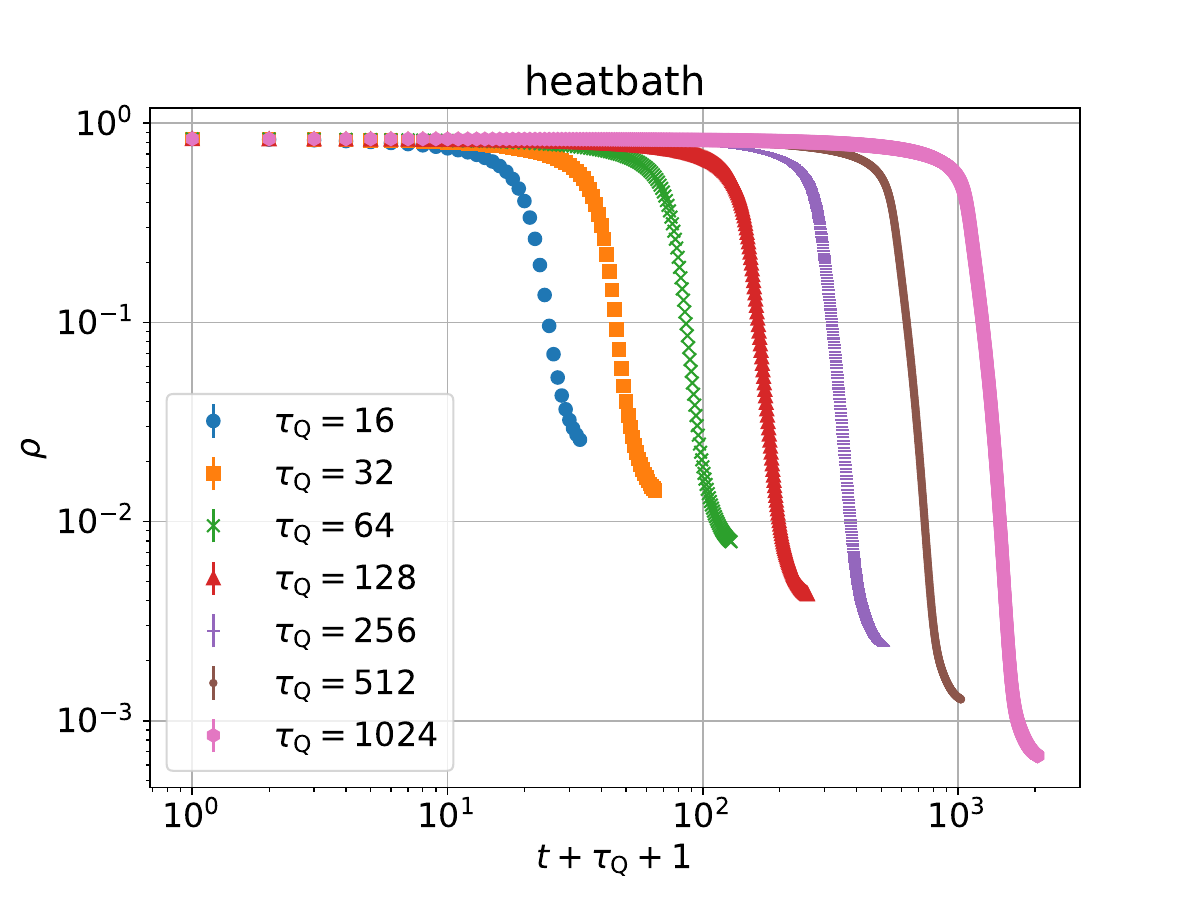}
\end{center}
\vspace*{-7mm}
\caption{The vortex density $\rho$
in a $96^{3}$ lattice under linear cooling, from
$T_{\rm i} = 2T_{\rm c}$ to $T_{\rm f} = 0$. The plots on the left
and on the right refer to the Metropolis and to the heatbath
algorithm, respectively.
The Markov time $t$ is given in units of lexicographic sweeps.}
\vspace*{-2mm}
\label{coolfig}
\end{figure}

Figure \ref{coolfig} shows the vortex density $\rho$, {\it i.e.}\
the fraction of all plaquettes that carry either a vortex or an
anti-vortex, on a $96^{3}$ lattice, during a cooling process which
is linear in the Markov time $t$ (in units of sweeps),
from $T_{\rm i} = 2T_{\rm c} \simeq 4.4$
down to $T_{\rm f} = 0$, according to eq.\ (\ref{lineartime}).
Of course, under rapid cooling --- with a short inverse
cooling rate $\tau_{\rm Q}$ --- more vortices remain.
Since the heatbath algorithm is more efficient than Metropolis,
less vortices persist after the same number of sweeps.

\begin{figure}[h!]
\vspace*{-5mm}
\begin{center}
\includegraphics[angle=0,width=.5\linewidth]{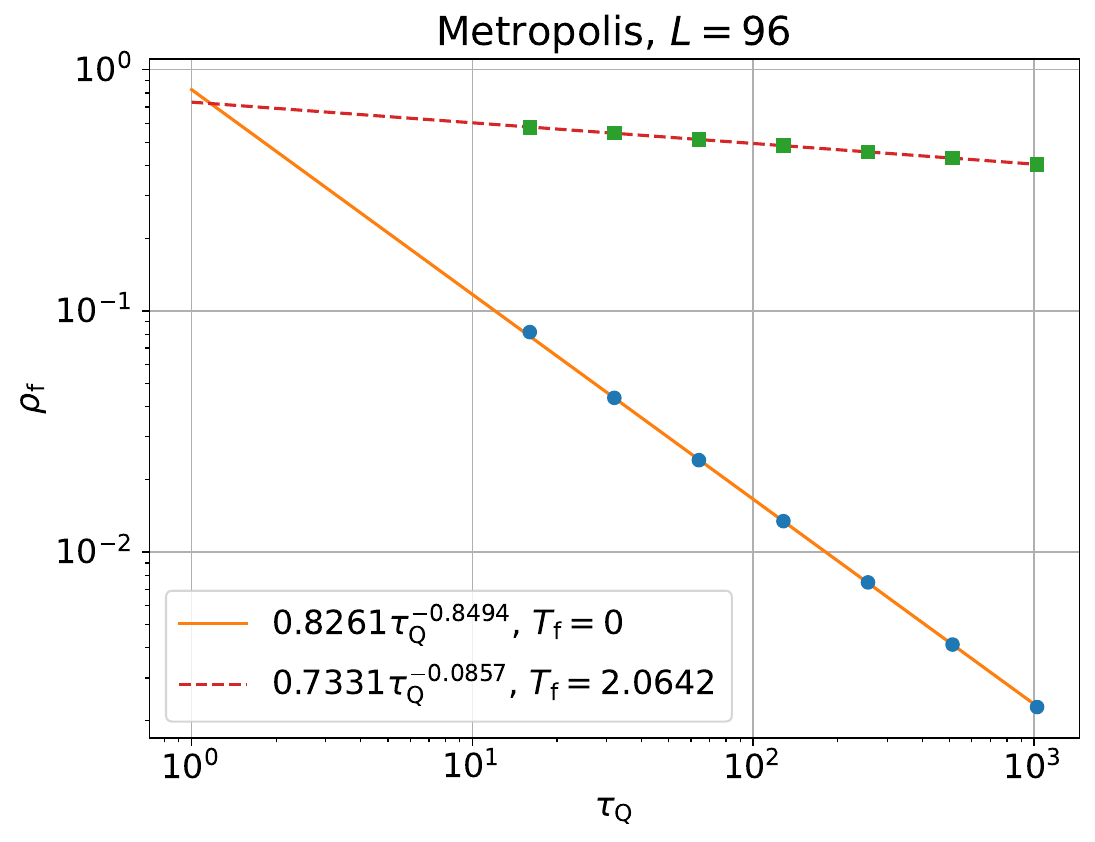}
\hspace*{-2mm}
\includegraphics[angle=0,width=.5\linewidth]{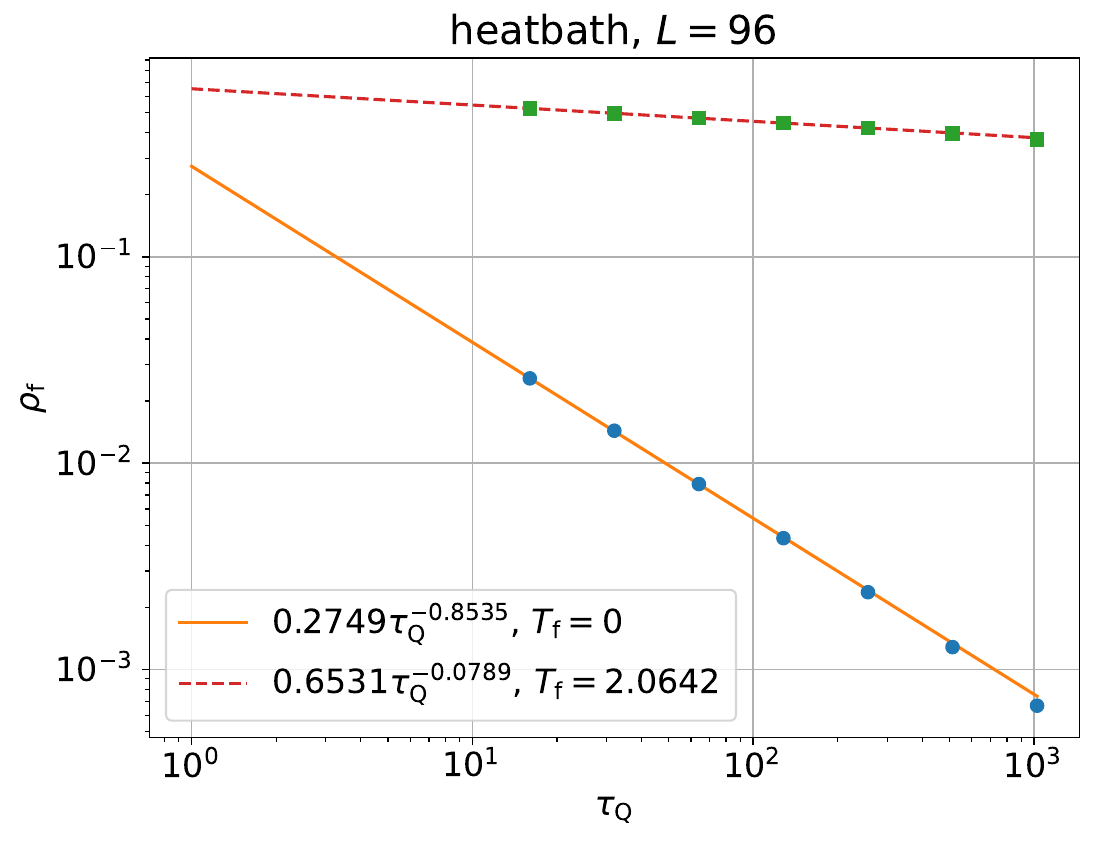} \\
\includegraphics[angle=0,width=.5\linewidth]{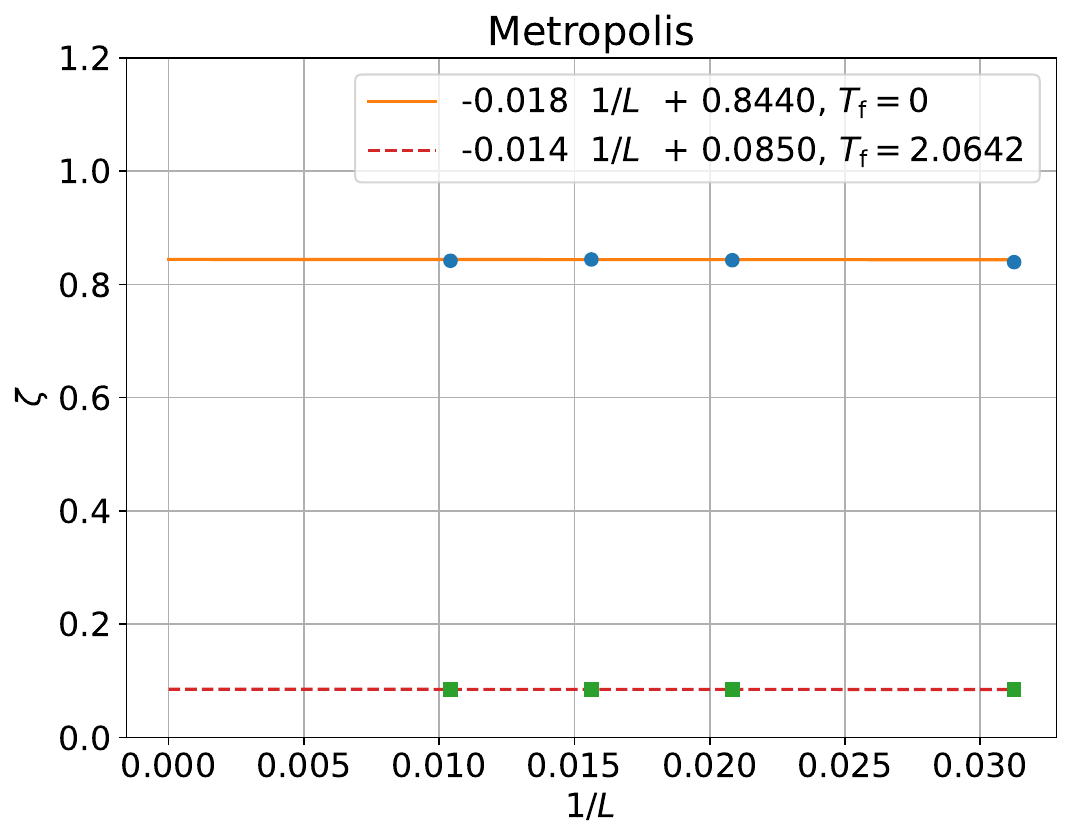}
\hspace*{-2mm}
\includegraphics[angle=0,width=.5\linewidth]{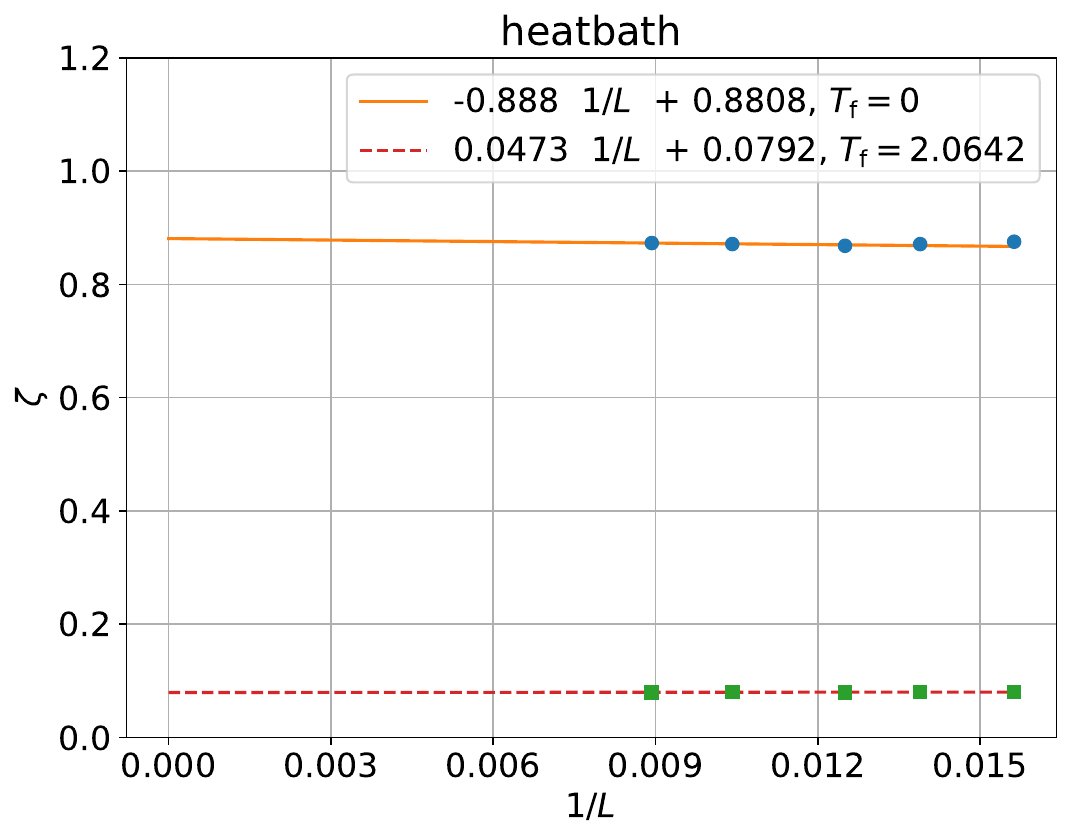} \\
\includegraphics[angle=0,width=.515\linewidth]{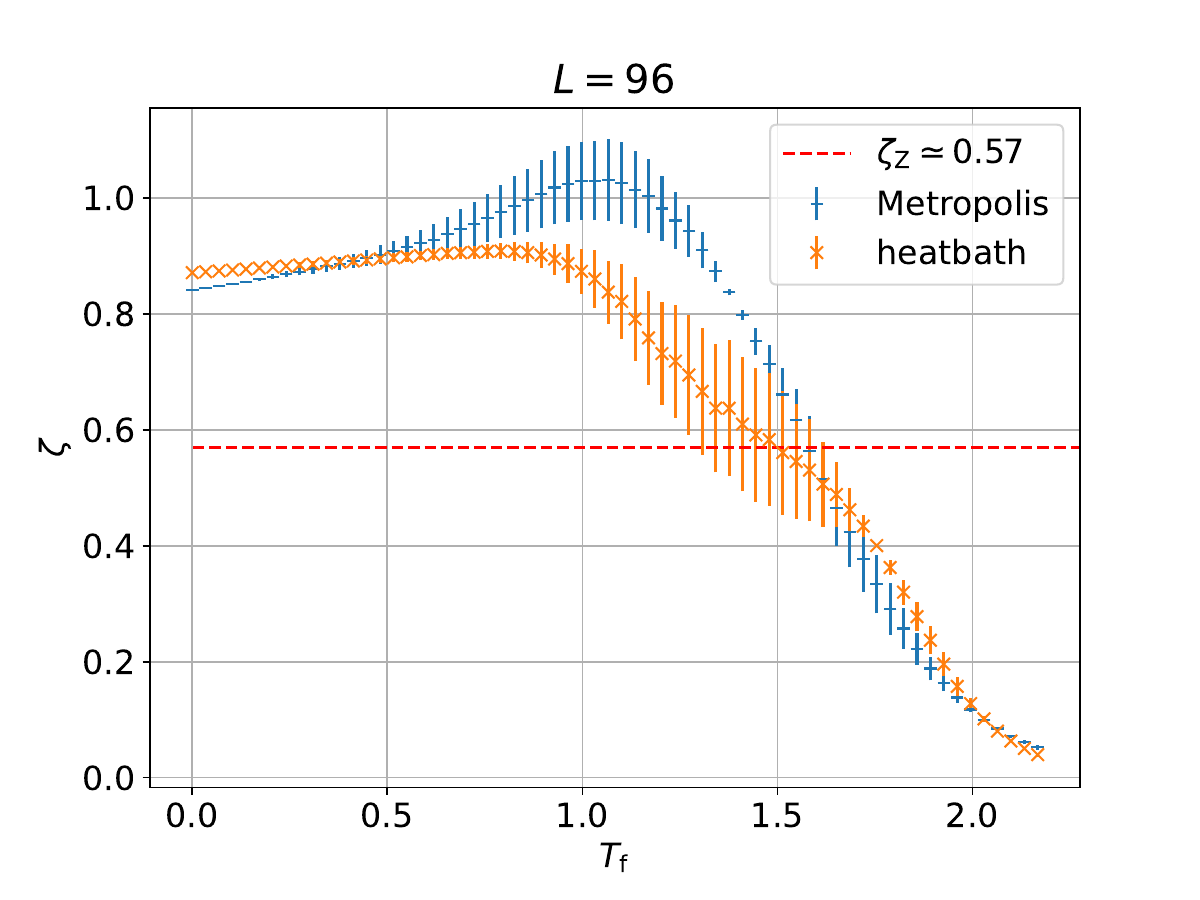}
\hspace*{-7mm}
\includegraphics[angle=0,width=.515\linewidth]{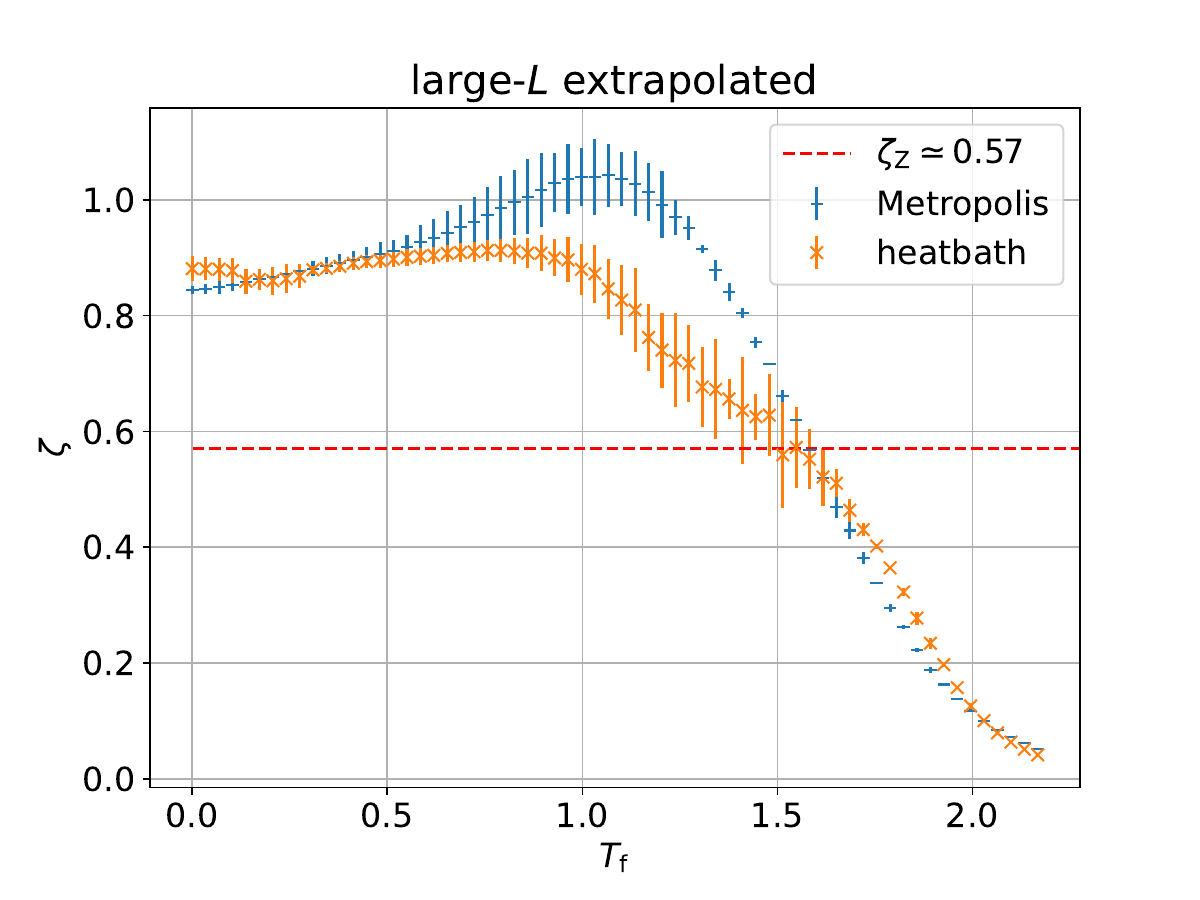}
\end{center}
\vspace*{-6mm}
\caption{Above: The persisting vortex density $\rho_{\rm f}$
at the end of a rapid cooling process down to two values of
$T_{\rm f} < T_{\rm c}$, in a lattice volume $96^{3}$. In each case,
$\rho_{\rm f}$ follows a power-law in $\tau_{\rm Q}$, in agreement
with eq.\ (\ref{rhoend}), $\rho_{\rm f} \propto \tau_{\rm Q}^{-\zeta}$.
Center: Large-$L$ extrapolations of the corresponding exponent
$\zeta$.
Bottom: Dependence of the $\zeta$-value on $T_{\rm f}$,
in a fixed lattice volume of $96^{3}$ (left) and large-$L$
extrapolated (right), compared to Zurek's prediction $\zeta_{\rm Z}$
in eq.\ (\ref{zetaZurek}).}
\vspace*{-10mm}
\label{powerfig}
\end{figure}

In Figure \ref{powerfig}, we first show the final vortex densities
$\rho_{\rm f}$, at the end of the cooling process, as a function
of $\tau_{\rm Q}$, for $T_{\rm f}=0$ and for $T_{\rm f}=2.0642$
(just below $T_{\rm c} \simeq 2.20$). The plots above confirm
the expected power-laws $\rho_{\rm f} \propto \tau_{\rm Q}^{-\zeta}$,
cf.\ eq.\ (\ref{rhoend}), which is observed for all $T_{\rm f} < T_{\rm c}$,
and for both algorithms.

The central plots illustrate the extrapolations of the
exponent $\zeta$, defined in eq.\ (\ref{rhoend}),
to the thermodynamic limit $L \to \infty$
(still referring to lattice volumes $L^{3}$).
The lower plots show how the values of $\zeta$ depend on $T_{\rm f}$,
at fixed $L$ and large-$L$ extrapolated.
Zurek's prediction (\ref{zetaZurek}) is attained for
specific final temperatures $T_{\rm f}$, but at this point this seems
accidental. Careful measurements of the auto-correlation time
--- to be identified with the relaxation time $\tau$ ---
will be required to check whether or not these values are related
to the cooling time $\hat t$ that Zurek refers to.

\vspace*{-3mm}
\section{Summary and outlook}
\vspace*{-2mm}

We presented simulation results or the 3d O(2) model as an
effective theory for the origin of (possible) cosmic strings
\cite{Kibble}.
The vortices form closed loops, like global cosmic strings.
We measured their mean length and density in thermal
equilibrium, as functions of the temperature.

We monitored the vortex density $\rho$ under rapid cooling,
across the critical temperature $T_{\rm c}$, which takes the system
out of equilibrium. Both with the Metropolis and with the heatbath
algorithm, the finally persisting density $\rho_{\rm f}$ follows a
power-law in the cooling rate, as predicted by Zurek \cite{Zurek85}.
The characteristic exponent $\zeta$ depends somewhat on the
algorithm and on the final temperature $T_{\rm f} < T_{\rm c}$.

A detailed study of the specific value of $T_{\rm f}$, at which Zurek
predicts the value of $\zeta_{\rm Z}$ in eq.\ (\ref{zetaZurek}),
remains to be performed. The
verification of his value depends on exhaustive measurements of
the auto-correlation time, which takes the role of the relaxation
time $\tau$.

\noindent
\ \\
{\bf Acknowledgments:}
We thank Jos\'{e} Armando P\'{e}rez Loera for his collaboration
in the framework of this project.
The simulations were performed on the cluster of the Instituto
de Ciencias Nucleares; we thank Luciano D\'{\i}az Gonz\'{a}lez
and Eduardo Murrieta Le\'{o}n for technical assistance.
This work was supported by UNAM-DGAPA through the PAPIIT project
IG100322 and by the Consejo Nacional de Humanidades, Ciencia
y Tecnolog\'{\i}a (CONAHCYT).

\end{document}